\documentclass[runningheads,a4paper]{llncs}

\usepackage{amssymb}
\usepackage{graphicx}

\usepackage{url}
\urldef{\mailsa}\path|{dhowe, cmpich}@springer.com|

\begin{document}

\mainmatter  

\title{Ontology Usage at ZFIN}
\titlerunning{Ontology Usage at ZFIN}

\author{Doug Howe and Christian Pich}
\authorrunning{Ontology Usage at ZFIN}
\institute{Zebrafish Model Organism Database,\\
5291 University of Oregon, Oregon, OR 97403, USA\\
\url{http://zfin.org}}

\toctitle{Ontology Usage at ZFIN}
\tocauthor{Authors' Instructions}
\maketitle

\begin{abstract}
The Zebrafish Model Organism Database (ZFIN) provides a Web resource of zebrafish genomic, genetic, developmental, and phenotypic data. Four different ontologies are currently used to annotate data to the most specific term available facilitating a better comparison between inter-species data. In addition, ontologies are used to help users find and cluster data more quickly without the need of knowing the exact technical name for a term.
\end{abstract}

\section{Introduction}

ZFIN is the model organism database for Danio rerio, the zebrafish and provides a centralized resource for zebrafish genomic, genetic, phenotypic, and developmental data. The ZFIN database contains highly integrated, manually curated information about genes, gene expression, mutant phenotypes and antibodies\cite{ZFIN}. Web-based search interfaces and tools allow viewing and analysis of the data, facilitate the understanding of gene function and regulation, and promote scientific discovery.

One significant obstacle when searching for annotated data is knowing the exact ontology term name to search with.  A curator may use one name to annotate a phenotype with a given anatomical structure while a user may know the same structure by a different name. To overcome this problem ontologies are created that use a definitive name for each entity and support the extensive use of synonyms. Such a dictionary is built as a directed acyclic graph (DAG) in which entities have one or more relationships to each other in a noncyclical pattern. DAGs provide a way to structure the various entities being modeled and lend themselves for various reasoning in regards to parent-child relationship questions.
Currently, ZFIN uses four different ontologies\cite{OBO} to annotate gene function, gene expression, and phenotypic information: Gene Ontology (GO)\cite{GO}, Zebrafish Anatomy Ontology (AO), Entity Quality Ontology (PATO) and the Spatial Ontology. Curators at ZFIN try to annotate to the most specific ontology term available.  This can make it hard for end users to find those annotations as they may not know the specific term name or they may wish to query with a more general term. Fortunately, due to the DAG-structure of the ontologies, annotations can be looked up by a higher level term name by performing an ontologically aware search that includes all subterms (children terms) if desired.  For example, a researcher is looking for genes that are expressed in the eye. The smart search returns all expression data that are annotated directly to the term eye and all it's child terms as well, e.g. retinal pigmented epithelium. In addition, ZFIN provides auto-completion as the user types a term name into a search entry field, i.e. a list of terms is provided that matches the users input. Synonymous names for all terms are included and help the user locate the best entity match.

\section{Current Challenges}
\begin{figure}
\includegraphics[width=14cm]{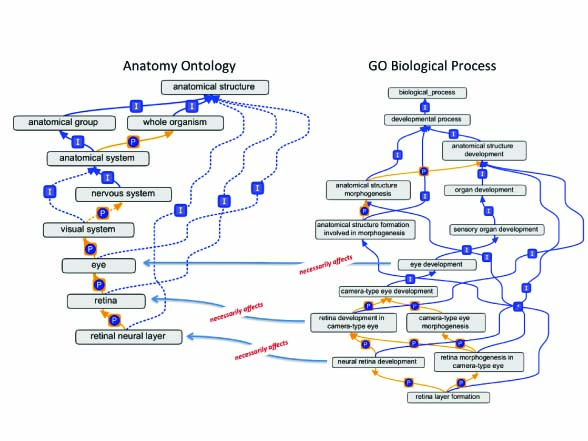}
\caption{Anatomy Ontology (left) and Gene Ontology (right) interrelationship.}
\label{fig:ontology-relationship}
\end{figure}

Ontologies are useful tools to capture and describe expression, phenotype, and gene functional data.  However ontologies are often incomplete and sometimes annotation of one observation can be made in several distinct ways.   Consider situations in which a term is not found in a given ontology but the term might be created as a cross-product of terms from two or more independent ontologies.  For example, if the term "fin development" was not currently in the GO, it could be emulated as a cross product of the term "fin" from the zebrafish AO and the GO term "development".  Currently, ZFIN accomplishes this through a technique of post-composition, annotation-time term composition in which a new entity is emulated as the intersection of two existing ontology terms. Including such post-composed terms in term lookups, annotation displays, download files, and web-based search results then becomes a more intricate problem. If a user queries for phenotypes affecting the anatomical structure  "actinotrichium", which is part of the fin, should an annotation with the post-composed term "fin":"development" be included in the results set?   If so, how and when is this association accomplished?  Even more difficult is following the proper logic when a phenotype annotation uses an existing GO term, like "neural retina development" and a user then makes a query for phenotypes that affect the "retina" by specifying the "retina" term from the zebrafish AO.  How should the zebrafish AO term "retina" logically and rigorously return phenotypes that are annotated with the GO term "neural retina development"?  There is no direct logical link (see Fig. \ref{fig:ontology-relationship}) other than a string match between the zebrafish AO term "retina" and the GO term "neural retina development". One way to remedy the missing links could be to create a separate
ontology that contains the relationships between the terms of the two
ontologies in question.  Work has already begun on this path in the case
of linking species specific anatomy ontologies and the biological
processes of the Gene Ontology\cite{REF}.

To answer this query extensive logical traversal of ontologies is necessary at the time the query is initiated or when data are indexed.  Extending the example further, consider a phenotype annotation involving the GO term "neural retina development".  If a researcher then makes a query for all phenotypes affecting the zebrafish eye by using the AO term "eye" in the ZFIN mutant search form, extensive logical reasoning must ensure to link the annotation using GO:"neural retina development" to the user query for phenotypes affecting the AO:"eye".  ZFIN is just beginning to explore this area of logical ontology traversal and we expect it will become an increasingly important aspect of data retrieval at ZFIN in the future.  We solicit your input on techniques to handle advanced logical ontology traversal, data linking, and data indexing strategies.

\section{Future Ontology-Driven Directions}

\begin{enumerate}
    \item Sophisticated logical reasoning to return correct and complete data sets regardless of how annotation was made
    \item Faceted data navigation interfaces
    \item Data-linked ontology browsing
    \item Incorporation of new ontologies (ChEBI, SO\cite{SO}, others)
\end{enumerate}

\end{document}